# Experimental investigations of local-time effect existence on laboratory scale and heterogeneity of space-time.

V.A. Panchelyuga[1], V.A. Kolombet[1], M.S. Panchelyuga[1] and S.E. Shnoll[1,2]

*Institute of Theoretical and Experimental Biophysics of RAS* (1)
*Lomonosov Moscow State University* (2)

*panvic333@yahoo.com, shnoll@iteb.ru*

The main subject of the work is experimental investigation of local-time effect existence on laboratory scale, which means longitudinal distances between locations of measurements from tens to one meter. Also short revue of our investigations of local-time effect existence for distances from 15 km to 500 m are presented. Besides investigations of the minimal spatial scale of local-time effect existence the paper presents investigations of the named effect for time domain. In this relation a structure of intervals distribution in neighborhood of local-time peak was studied and splitting of the peak was found out. Further investigations shows second order splitting of local-time peak. From this result arise a supposition that space-time heterogeneity, which following from local-time effect existence probably has fractal character. Obtained results lead to conclusion about sharp anisotropy of space-time.

## 1. Introduction.

Our previous works [1 – 4] give detailed description of macroscopic fluctuations phenomena, which consists in regular changes of fine structure of histograms shapes built on the base of short samples of time series of fluctuations in different process of any nature – from biochemical reactions and noises in gravitational antenna to fluctuations in $\alpha$-decay rate. From the fact that fine structure of histograms, which is main object of macroscopic fluctuations phenomena investigations, doesn't depend on qualitative nature of fluctuating process follows a conclusion that the fine structure can be caused only by such common factor as space-time heterogeneity. Consequently, macroscopic fluctuations phenomena can be determined by gravitational interaction or as shown in [5, 6] by gravitational wave influence.

    Present work was carried out as further investigations of macroscopic fluctuations phenomena. The local time effect, which is the main subject of this paper, is synchronous in local time appearance of pairs of histograms with similar fine structure constructed on the base of measurements of fluctuations in processes of different nature fulfilled in different geographical locations. The effect points out on the dependence of fine structure of the histograms on the Earth rotations around its axis and around the Sun.

    The local time effect is closely connected with space-time heterogeneity. In other words, this effect is possible only if experimental setup, consisting of pair of spaced sources of fluctuations moving through heterogeneous space. It is obvious that for the case of homogeneous space the named effect don't exist. Existence of local-time effect for some space-time scale can be considered as evidence of space-time heterogeneity, which corresponds to this scale. Dependence of local-time effect on the local time or longitudinal difference between places of measurements leads to conclusion that space heterogeneity has axial symmetry.

    The existence of local time effect was studied for different distances between places of measurement from hundred kilometers up to highest possible on the Earth distances (~ 15000 km). The goal of the present work is the investigations of the existence of the effect for distances between places of measurements from tens meters up to one meter. Such distances in title of the paper are named as 'laboratory scale'.

## 2. Experimental investigations of local-time effect existence for longitudinal distances between places of measurements from 15 km to 500 m.

The main problem of experimental investigations of local-time effect at the small space distances is resolution enhancement of the macroscopic fluctuations method, which defined by histograms duration. As a rule all investigations of local-time effect were carried out by using α-decay rate fluctuations of $^{239}$Pu sources. Histograms duration in this case are one minute. But such source of fluctuations becomes uselessness for distances in tens of kilometers or less when histograms duration must be about one second or less. By this reason in work [7-8] we refuse α-decay sources of fluctuations. As such a source was chosen noise generated by semiconductor diode. Used diodes gives noise signal with frequency band up to tens of megahertz and because of this satisfies the requirements of present investigations.

To check convenience of selected noise source for local-time effect investigations it was tested on distances for which existence of the effect was proved by using of α-decay sources of fluctuations [7]. This work was shown appropriateness of semiconductor noise diode for studies of local-time effect.

Below we presents short description of our experiments dedicated to investigation of local-time effect for longitudinal distances between locations of measurements from 15 km up to 500 m. The first experiment is study of the local-time effect existence for longitudinal distance between locations of measurements equal 15 km, the second one for set of distances from 6 km to 500 m. More detailed description of this experiments is given in [8].

In first experiment series of synchronous measurements were carried out in Pushchino (Lat. 54°50.037′ North, Lon. 37°37.589′ East) and Bolshevik (Lat. 54°54.165′ North, Lon. 37°21.910′ East). Longitudinal difference α between places of measurements was α = 15.679′. This value of α corresponds to difference of local time $\Delta t$ = 62.7 sec and longitudinal distance $\Delta l$, equal $\Delta l$=15 km.

To study local-time effect in Pushchino and Bolshevik were obtained 10-minute time series by digitizing fluctuations from noise generators with sampling frequency equal to 44100 Hz. From this initial time series with three different steps equal 735, 147, and 14 points were extracted single measurements and obtained three time series with equivalent frequency equal 60 Hz, 300 Hz and 3150 Hz. On the base of this time series in a standard way [1-3] on the base of 60-points sample length for first and second time series and 63-points sample length for third time series were constructed three sets consisting of histograms with duration 1 sec, 0.2 sec and 0.02 sec.

Fig. 1 presents intervals distribution obtained after expert comparisons of 1-sec histogram sets. The distribution has a peak, which corresponds to time interval equal to 63±1 sec, which with good accuracy corresponds to local time difference $\Delta t$ = 62.7 sec between places of measurements.

Local time peaks ordinary obtained on the interval distributions are very sharp and consist of 1-2 histograms [1-3] i.e. is practically structureless. Peak on the Fig. 1 *a*) also can be considered as structureless. This reason leads us to the problem of further investigating of its structure.

The fact that all sets of histograms were obtained on the base of the same initial time series from one hand enables enhancement of time resolution of the method of investigation and from the other – eliminates necessity of very precise and expensive synchronization of spaced measurements. Intervals distribution obtained for 1-sec histograms set allows using information about location of local-time peak alignment of time series. The alignment makes possible using set of histograms of the next order of smallness.

Using of 0.2-sec histograms set increase resolution in five times and allows more detailed investigations of local-time peak structure and its position on the time axis. Since the positions of the peak on the 1-sec intervals distribution (Fig. 1) are known it is possible to select their neighborhood by means of 60 sec relative shift of initial time series and prepare after this 0.2-sec histograms set for further expert comparison.

Intervals distribution obtained in result of expert comparisons for 0.2-sec histograms set is presented on Fig. 1*b*). One can see that maximum similarity of histograms shape is observed for pairs of histograms separated by interval in 63±0.2 sec. This value is the same as for 1-sec histograms intervals distribution, but in latter case it is defined with accuracy in 0.2 sec.

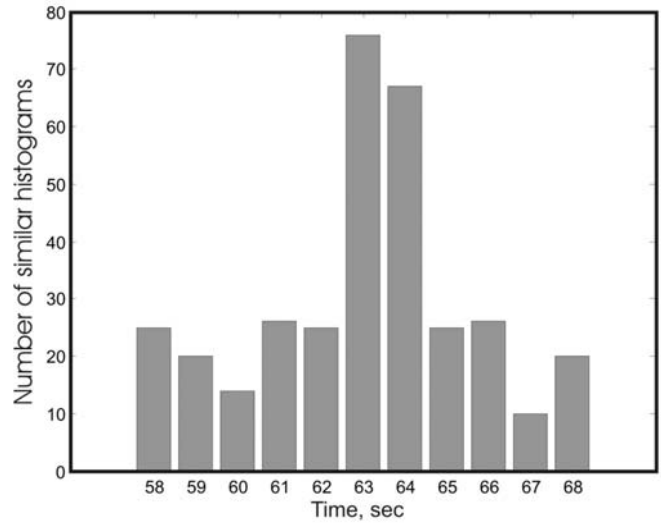

*a*)

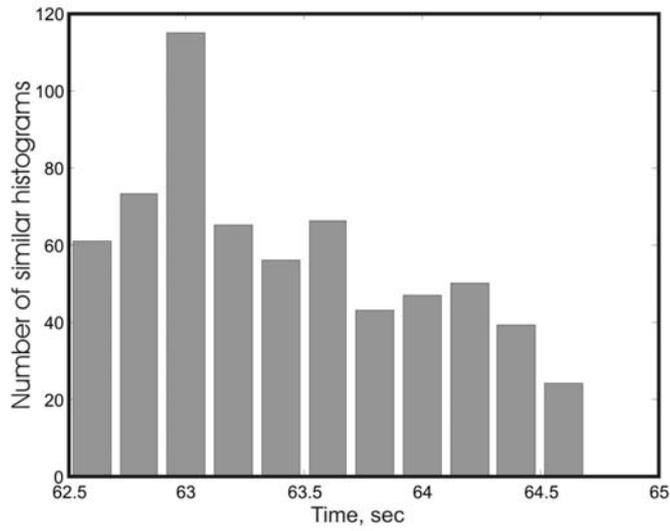

*b*)

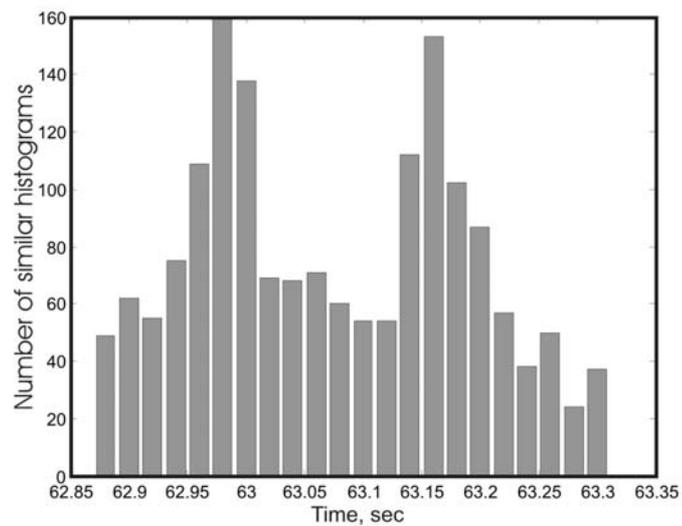

*c*)

Fig. 1. Intervals distributions obtained after expert comparisons of 1-sec (*a*), 0.2-sec (*b*), and 0.02-sec (*c*) histogram sets. Y-axis presents number of histograms, which were found similar; X-axis – time interval between pairs of histograms, sec.

It's easy to see from intervals distribution, Fig. 1*b*), that after 5-times enhancement of resolution, the distribution has single sharp peak again. So, change of time scale in this case doesn't lead to change of intervals distribution. This means that we must enhance time resolution again to study the local time peak structure. We can do this by using of 0.02-sec histograms set.

Intervals distribution for case of 0.02-sec histograms is presented on the Fig. 1*c*). Unlike to intervals distributions on the Fig. 1*a*) and Fig. 1*b*) distribution on the Fig. 1*c*) consists of two distinct peaks. The first peak corresponds to local time difference equal 62.98±0.02 sec, the second one to 63.16±0.02 sec. The difference between the peaks is $\Delta t' = 0.18 \pm 0.02$ sec.

Splitting of local-time peak on the Fig 1 *c*) is similar to splitting of daily period on two peaks with periods, which equal to solar and sidereal days [9-11]. This result will be considered in the next section.

Above described experiment demonstrates the existence of local-time effect for longitudinal distance between locations of measurements in 15 km and splitting of local-time peak corresponding to that distance. It is natural to investigate the question: which is the minimal distance of local time effect existence? Next step in this direction is the second experiment presented below.

In this experiment two measurement systems were used: stationary and mobile one. Four series of measurements were carried out. Longitudinal difference of locations of stationary and mobile measurement systems was 6 km, 3.9 km, 1.6 km and 500 m. Method of experimental data processing was the same as for first experiment. Was found that for every of above presented distances local-time effect exists and the local-time peak splitting can be observed.

## 3. Second-order splitting of local-time peak. Preliminary results.

Four-minute splitting of daily period of repetition of histograms shape on solar and stellar sub-periods was reported in [3]. In sited paper the phenomena is considered as an evidence of existence of two preferential directions: to the Sun and to the coelosphere. Really after time interval equal 1436 min the Earth makes one complete revolution and measurement system plane has the same direction in the space as one stellar day before. After four minutes from this moment measurement system plane will be directed to the Sun. This is the cause of solar-day period – 1440 min.

Let us suppose that splitting described in the present paper has the same nature as splitting of daily period. Then from daily period splitting $\Delta T$, which equal $\Delta T = 4\,min$ its possible to obtain proportionality coefficient $k$:

(1) $$k = \frac{240\,\text{sec}}{86400\,\text{sec}} \approx 2.78 \cdot 10^{-3}.$$

Longitudinal difference between places of measurements presented in second section is $\Delta t = 62.7$ sec and we can calculate splitting of local-time peak for this value of $\Delta t$:

(2) $$\Delta t' = k\Delta t = 62.7 \times 2.78 \cdot 10^{-3} \approx 0.17\,\text{sec}.$$

As it is easy to see from Fig. 1*c*) splitting of local-time peak is equal to 0.18±0.02 sec. This value agrees with estimation (2). Values of splitting of the local-time peak, which was found for mobile experiment, also are in good agreement with values obtained by the help of formula (2).

This result allows us to consider sub-peaks of local-time peak as stellar and solar and suppose that in this case the cause of splitting can be the same as for daily-period splitting. But the question about local-time peak structure remains open.

In order to further investigations of the local-time peak structure was carrying out the experiment on synchronous spaced measurements in Rostov-on-Don (Lat. 47°13.85′ North, Lon. 39°44.05′ East) and Bolshevik (Lat. 54°54.16′ North, Lon. 37°21.91′ East). Local-time difference for locations of measurements is $\Delta t$ = 568.56 sec. Value of local-time peak splitting, according to (2), is $\Delta t' = 1.58$ sec. Method of experimental data processing was the same as described in second section.

On the fig. 2 in form of diagram summa of all expert comparisons results are presented. For considered case we avoid presentations of our results in the form of intervals distributions, like on fig 1, because of it multiplicity (about of ten graphs).

Diagram on the fig. 2 consists of four lines. At the leftmost side of every line is presented duration of single histogram in the four sets of histograms, which was prepared for expert comparisons. So, we have four sets consisting of 1-sec, 0.2-sec, 0.0286-sec and 1.36 msec histograms. Rectangle in the first line schematically shows local-time peak, obtained as result of expert comparisons of 1-sec histograms set. Taking into account synchronization error (about one second), expert result is 567±2 sec. This value is in agreements with calculated longitudinal difference of local time Δ*t* = 568.56 sec (everywhere on the diagram, fig. 2, calculated values are given in parenthesis).

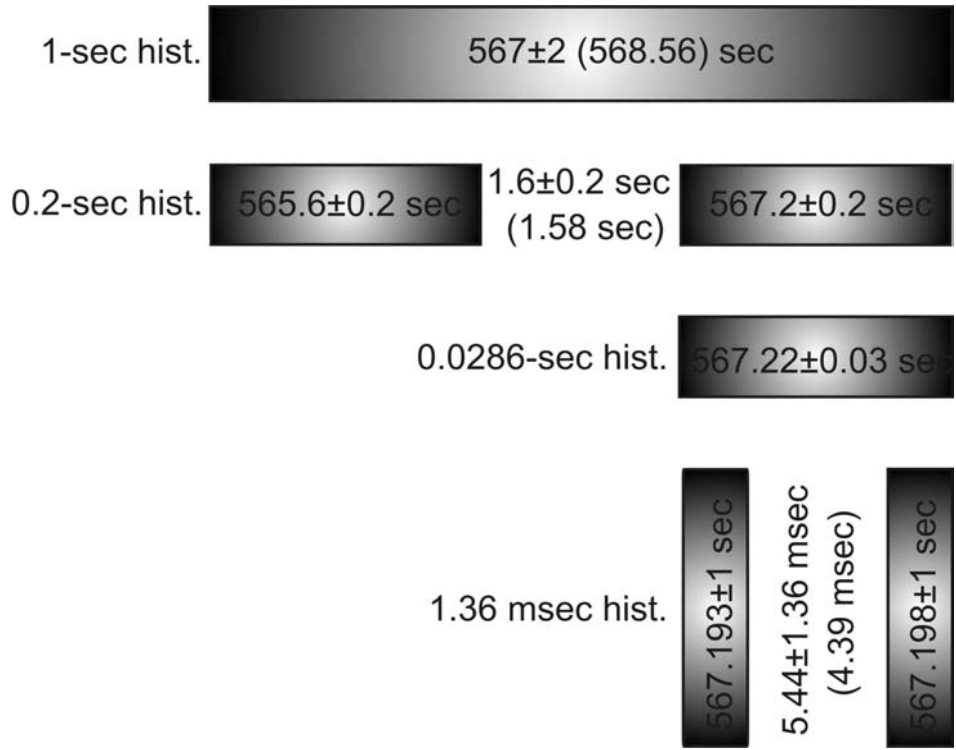

Fig. 2. Local-time peak splitting obtained in the experiment on synchronous spaced measurements of fluctuations of pair of semiconductor noise generators, which was carried out in Rostov-on-Don and Bolshevik.

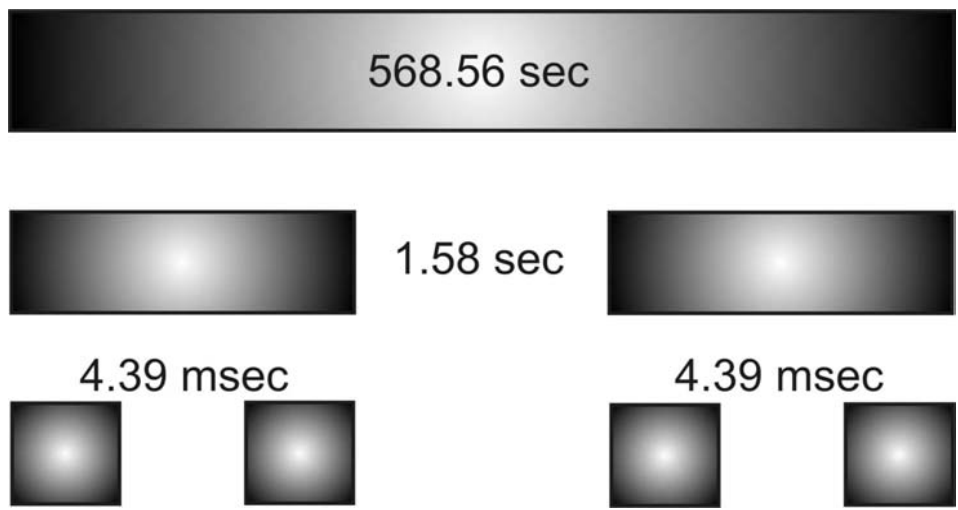

Fig. 3. Expected structure of local-time peak splitting for experiment on synchronous spaced measurements in Rostov-on-Don and Bolshevik, which was calculated on the base of formula (4).

Second line on the fig. 2 presents results for 0.2-sec histograms set. Values in rectangles show sidereal and solar subpeaks of the local-time peak. Value between the rectangles gives splitting of local-time peak. Experimentally obtained splitting value is 1.6±0.2 sec. This value is in good agreement with value calculated on the base of formula (2).

Third and fourth lines of diagram present further investigation of local-time peak structure. In the third line we have result of expert comparisons of 0.0286-sec histograms set for intervals, which constitute closest neighborhood of 567.2±0.2-sec peak. Using of 0.0286-sec histograms set increase resolution almost in ten times and defines peak position on the intervals distribution at 567.22±0.03 sec. Obtained peak is structureless. Further increasing of resolution allows 1.36-msec histograms set, presented in fourth line. In this case resolution enhancement discover splitting of 567.22±0.03 sec peak.

Splitting, presented in last line of the diagram can be named as second-order splitting. It can be calculated using first-order splitting $\Delta t' = 1.58$ sec by the analogue of formula (2):

(3) $$\Delta t'' = k\Delta t'.$$

As easy to see from (3) and from diagram on fig. 2, for second-order splitting $\Delta t''$ value of first-order splitting $\Delta t'$ plays the same role as local-time value $\Delta t$ for $\Delta t'$. Numerical calculations using (3) gives $\Delta t'' = 4.39$ msec. This estimation of $\Delta t''$ is in good agreement with experimentally obtained splitting value 5.44±1.36 msec.

Experimental evidence of existence of second-order splitting leads us to supposition about possibility of *n*-order splitting. As easy to see from (2) and (3), *n*-order splitting value $\Delta t^n$ can be obtained in the following way:

(4) $$\Delta t^n = k^n \Delta t.$$

Fig. 3 presents idealized structure of local-time peak splitting for considered experiment, which was calculated on the base of formula (4). Unlike to fig. 2 structure of local-time peak splitting on the fig. 3 is symmetrical. Studies of possible splitting of 565.6±0.2 sec peak is our immediate task. At the time, results presented on the fig. 2 can be considered as preliminary.

## 4. Experimental investigations of local-time effect existence for longitudinal distances between places of measurements from 12 m to 1 m.

Experiments described in two previous sections demonstrate existence of local-time effect for longitudinal distance between locations of measurements in 500 m and existence of second-order splitting of local-time peak. Next step in our investigations is study of local-time effect existence on the laboratory scale.

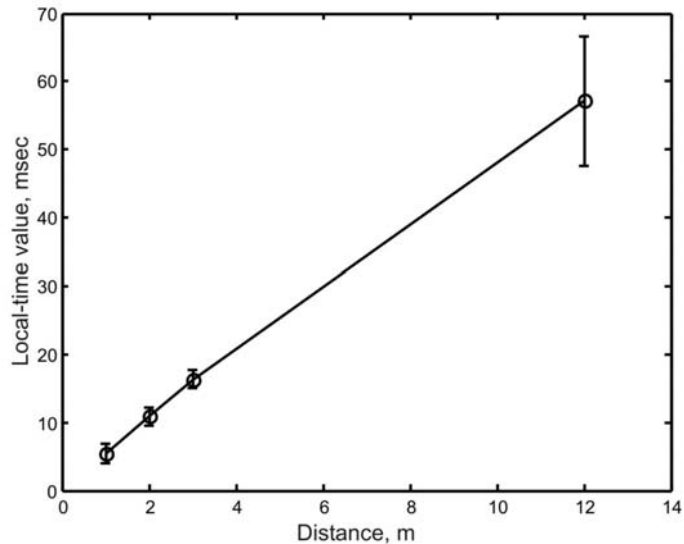

Fig. 4. Values of local time shift as function of distance between two sources of fluctuations. The graph presents results of investigations of local-time effect existence for distances 12 m, 3m, 2m, and 1 m.

The main difference of local-time effect investigations on laboratory scale from experiments described above is absence of special synchronization system. In laboratory case experimental setup consists of two synchronous data acquisition channels and two spaced noise generators, which symmetrically connected to it. In role of data acquisition system was used LeCroy WJ322 digital storage oscilloscope. Standard record length of the oscilloscope consists of 500 kpts per channel. This allows obtaining two synchronous sets of 50-points histograms. Maximum length of every set is 10000 histograms.

Fig. 4 presents values of local time shift as function of distance between two noise generators. The graph presents results of investigations of local-time effect existence for distances in 12 m, 3m, 2m, and 1 m. Local-time values was found with accuracy in 9.52 msec for 12-meters experiment and with accuracy in 1.36 msec for 1 m, 2 m, and 3 m experiments.

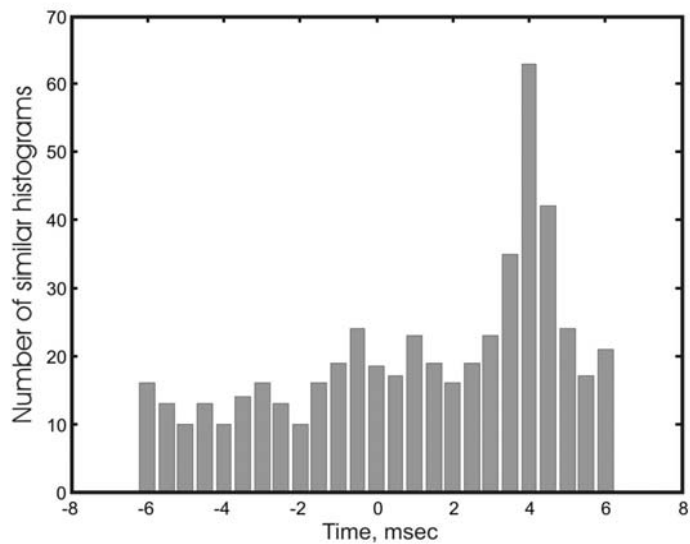

Fig. 5. Example of intervals distribution for longitudinal distance between two spaced sources of fluctuations in one meter. Single histogram duration – 0.5 msec.

An example of intervals distribution for longitudinal distance between two noise generators in one meter is presented on fig. 5. The intervals distribution was obtained on the base of 0.5-msec histograms set. By using of Earth equatorial radius value (6378245 m) and latitude of place of measurements (54°50.0.37′) is possible to estimate local-time difference for longitudinal distance in one meter. Estimated value is 3.7 msec. As easy to see from fig. 5, experimentally obtained value of local-time peak is 4±0.5 msec. This experimental value is in good agreement with theoretical one.

Results of our investigations for laboratory scale, which are presented in this section, confirm local-time effect existence for distances up to one meter. So, we can state that local-time effect exists for distances from thousands kilometers to one meter. This is equivalent to statement that space heterogeneity can be observed up to one-meter scale.

## 5. Discussions.

Local-time effect as pointed in [1], is linked to rotatory movement of Earth. The simplest explanation of the fact can be following. Due to the rotatory movement of the Earth after time $\Delta t$ measurement system No. 2 appears in the same places where was system No. 1 before. The same places cause the same shape of fine structure of histograms. Actually such explanation is incorrect because of orbital motion of Earth, which noticeably exceeds rotatory movements. Therefore measurement system No. 2 cannot appear in the same places where was system No. 1. But if we consider two directions defined by center of Earth and two points were we conduct spaced measurement, then after time $\Delta t$ measurement system No. 2 take the same directions in the space as system No. 1 before. From this it follows that similarity of histograms shapes in some way is connected with the same space directions. This supposition also agrees with experimental results presented in [12-13].

Speaking about preferential directions we implicitly supposed that measurement system is directional and because of this can resolve these directions. Such supposition is quite reasonable for the case of daily period splitting but for splitting of local-time peak observed on the one-meter scale becomes utterly problematic because an angle, which must be resolved by the measurement system, is neglible. Most likely that in this case we deal with space-time structure, which in some way are connected with preferential directions to the Sun and coelosphere. Second-order splitting of local-time peak also can be considered as an argument confirming this supposition. Apparently we can speak about sharp anisotropy of near-earth space-time. Existence of local-time effect leads us to supposition that this anisotropy is axial-symmetric.

Authors grateful to Dr. Hartmut Muller, V.P. Tikhonov and M.N. Kondrashova for valuable discussions and financial support. Special thanks to our colleagues O.A. Mornev, R.V. Polozov, T.A. Zenchenko, K.I. Zenchenko and D.P. Kharakoz.